\documentclass[10pt,a4paper]{article}
\usepackage[latin1]{inputenc}
\usepackage{amsmath}
\usepackage{amsfonts}
\usepackage{amssymb}
\usepackage{makeidx}
\usepackage{graphicx}
\usepackage{braket}
\usepackage{bm}
\usepackage{color}
\usepackage{hyperref}
\usepackage{dsfont}
\usepackage{enumerate} 
\usepackage{float} 
\usepackage{mathrsfs}
\usepackage{authblk}

\title{Dynamics of a composite quantum bouncing ball}
\author[1]{R. Chatterjee \thanks{riddhi.chatterjee@bose.res.in}}
\affil[1]{S. N. Bose National Centre For Basic Sciences, Block JD, Sector- III, Salt Lake, Kolkata- 700106}
\date{}
\begin{document}
\maketitle

\begin{abstract}
I construct the state of a quantum particle with internal degree of freedom bouncing on a perfect reflector in presence of gravity. I predict from the result that revival of position expectation is possible for such system.
\end{abstract}

\section{Introduction}
Atomic cloud bouncing on a perfectly reflecting surface(i.e mirror) was first experimentally demonstrated by Aminoff \textit{et al.} \cite{cloud}.
Quantum bouncing ball was analytically studied by J. Gea-Banacloche \cite{banac}. In the same time and afterwards it has been practically realised using Bose Einstein condensates \cite{bec1}, ultracold neutrons \cite{ucn} and found huge application especially in the field of non Newtonian interaction, gravimetry, dark energy \cite{appl} and many more. \par

It has been shown by Brukner \textit{et. al.} Centre of mass motion of a composite particle (particle with internal degree of freedom) is influenced by its internal degree of freedom(IDOF) due to contribution of IDOF into mass energy of the system. As a result mass becomes a dynamical variable resulting superposition of mass at low energy regime. These formulations lead to interesting phenomena such as decoherence due to gravitational time dilation and quantum violation of Einstein's equivalence principle \cite{bruk}. Though these formulations has resulted some controversy and alternative interpretations \cite{pang, hu, dio}.\par

In this article we will construct the state of a composite quantum bouncing ball.

\section{State of a composite quantum bouncing ball}

Hamiltonian of a free particle with internal degree of freedom in presence of gravtational field in laboratory reference frame is given by \cite{bruk} \begin{equation}
\hat{H}= \hat{M}c^2 + \dfrac{\hat{P}^2}{2\hat{M}} + \hat{M}g\hat{z}
\end{equation}
where $\hat{M} = m \hat{\mathbb{I}}_{int} + \hat{H}_{int}/c^2$, $\hat{\mathbb{I}}_{int}$ is the identity operator in the internal state space of the particle, $\hat{H}_{int}$
is internal Hamiltonian of the particle. Assuming lowest eigenvalue of $\hat{H}_{int}$ to be zero, $mc^2$ is the mass energy of the particle at ground state. $\hat{H}$ is valid upto first order correction of $\hat{H}_{int}/mc^2$ and hence the Hamiltonian can be written as $\hat{H} = \hat{H}_0 + \hat{H}_1$ where, \begin{equation}
\begin{split}
\hat{H}_0 = mc^2 + \dfrac{\hat{P}^2}{2m} + mg\hat{z} +\hat{H}_{int} \\ 
\hat{H}_1 =  \dfrac{\hat{H}_{int}}{mc^2} \Big(- \dfrac{\hat{P}^2}{2m} + mg\hat{z} \Big)
\end{split}
\end{equation}
To find the state of quantum bouncing ball we need to solve above Hamiltonian with boundary condition -- state $\psi = 0$ at $z = 0$ (because of the mirror). Whole problem is treated as one dimensional.\par

\subsection{Solution of $H_0$:}

Solution of Schr\"odinger equation for Hamiltonian $\Big( \dfrac{\hat{P}^2}{2m} + mg\hat{z} \Big)$ with above mentioned boundary condition is \cite{banac} $\psi_n = N_n Ai\Big( k \Big( z- \dfrac{E_n}{mg}\Big)\Big) = N_n Ai(kz -\alpha_n)$. Here $n$ corresponds $n^{th}$ eigenstate, $\{\alpha_n\}$ are zeroes of airy function, $k = (2m^2g/\hbar^2)^{1/3}$, $N_n$ is the normalisation constant. Eigenvalue $E_n = (mg\alpha_n/k)$.\par

So, if $\{\ket{i}\}$ is the eigenbasis for $\hat{H}_{int}$ with eigenvalues $\{E_i\}$ then eigenvalue equation for $\hat{H}_0$ is \begin{equation}
\hat{H}_0 \ket{\psi_n , i} = (mc^2 + E_n + E_i) \ket{\psi_n , i}
\end{equation}

\subsection{Correction for $H_1$:}

Here first order perturbation theory is applied. $\hat{H} = \hat{H}_0 + \hat{H}_1 = \hat{H}_0 + \lambda \hat{V}$ where $\lambda = \dfrac{1}{m c^2} \ll 1$ and  $\hat{V} = \hat{H}_{int} \Big(- \dfrac{\hat{P}^2}{2m} + mg\hat{z} \Big)$. Straightforward calculation gives --

\subsubsection*{Correction in energy:}

\begin{equation}
E^{(1)}_{ni} = \braket{\psi_n,i^{(0)}\mid \hat{H}_1\mid \psi_n,i^{(0)}} = \dfrac{E_i E_n}{mc^2}
\end{equation}

\subsubsection*{Correction in eigenstates:}

\begin{equation}
\begin{split}
\ket{\psi_n,i^{(1)}} = \sum_{\lbrace m,j\rbrace \neq \lbrace n,i\rbrace} \dfrac{\braket{\psi_m,j^{(0)}\mid \hat{H}_1\mid \psi_n,i^{(0)}}}{(E^{(0)}_{ni}) - E^{(0)}_{mj}}  \ket{\psi_m,j^{(0)}} \\
= \dfrac{E_i}{mc^2} \sum_{m(\neq n)}^{\infty} \dfrac{(-)^{m-n+2}2}{(E_m -E_n)^3} \ket{\psi_m,i^{(0)}}
\end{split}
\end{equation}

\section{Time development:}

Let initially particle is in a superposition state $\ket{\phi} = \sum_n c_n \ket{\psi_n} \otimes \sum_i a_i \ket{i}$. Density matrix of the particle after time $t$ \begin{equation}
\rho = \rho_{cm}(0) \otimes \rho_{int}
\end{equation}
After time t, \begin{equation}
\rho_t = e^{-\dfrac{it\hat{H}}{\hbar}} \cdot \rho \cdot e^{\dfrac{it\hat{H}}{\hbar}}
\end{equation}

Tracing out internal degree of freedom we get the reduced density matrix. If we assume IDOF(s) are in a thermal state with temperature $T$ \begin{equation} \label{red}
\rho_{red} = \sum_{i} \dfrac{e^{-\beta E_i}}{Z}\cdot e^{-\dfrac{itH(E_i)}{\hbar}} \cdot \rho_{cm}(0) \cdot e^{\dfrac{itH(E_i)}{\hbar}}
\end{equation}

where $H(E_i) = \dfrac{\hat{P}^2}{2(m+\dfrac{E_i}{c^2})} + (m+\dfrac{E_i}{c^2})g\hat{z}$. $Z$ is the partition function, $\beta = (k_B T)^{-1}$, $k_B$ is Boltzmann constant. Evaluating equation (\eqref{red}) we get \begin{equation}
\rho_{red} = \sum_{m,n,i} \dfrac{e^{-\beta E_i}}{Z}\cdot e^{-\Big(\dfrac{igt(\alpha_m - \alpha_n)(m+ \dfrac{E_i}{c^2})}{k_i \hbar} \Big)} c_m c_n^* \ket{\psi_m}\bra{\psi_n}
\end{equation}

Where $k_i =\Big( \dfrac{2(m + \dfrac{E_i}{c^2})^2g}{\hbar}\Big)^{1/3}$. Since $1/(mc^2)\ll 1$, we can write above expression upto first order in $1/(mc^2)$, as-- \begin{equation}
\rho_{red} = \sum_{m,n,i} \dfrac{e^{-\beta E_i}}{Z}\cdot e^{-\Big(\dfrac{igt(\alpha_m - \alpha_n)(m+ \dfrac{E_i}{3c^2})}{k \hbar} \Big)} c_m c_n^* \ket{\psi_m}\bra{\psi_n}
\end{equation}

Expanding around average energy $\bar{E}$ upto second order we get \begin{equation}
\rho_{cm} = \sum_{m,n} c_m c_n^* e^{-\Big(\dfrac{igt(\alpha_m - \alpha_n)(m+ \dfrac{\bar{E}}{3c^2})}{k \hbar} \Big)} e^{-\delta} \ket{\psi_m}\bra{\psi_n}
\end{equation}
where $\delta = \dfrac{\left\langle \Delta E^2\right\rangle t^2 g^2 (\alpha_m - \alpha_n)^2}{9 \hbar^2 c^4 k^2}$, $\Delta E$ is the fluctuation. Now $\left\langle \Delta E^2\right\rangle \propto T^2$ where $T$ is temperature associated to IDOF. Usually the factor $e^{-\delta}$ will suppress the off diagonal term of reduced density matrix. At low temperature $e^{-\delta} \rightarrow 1$. So, quantum state at low temperature is basically like a particle without any degree of freedom with an additional factor $\bar{E}/3c^2$ added with mass in the numerator.

\section{Conclusion:}

So, at low temperature we will observe usual collapse and revival of position expectation. Next we will study application of above formulation.

\end{document}